# Self-corrected sensors based on atomic absorption spectroscopy for atom flux measurements in molecular beam epitaxy*


Y. Du,[1,a)] T. C. Droubay,[2] A. V. Liyu,[1] G. Li,[3] and S. A. Chambers[2,b)]

[1]*Environmental Molecular Sciences Laboratory, Pacific Northwest National Laboratory, Richland, Washington 99352, USA*

[2]*Fundamental & Computational Sciences Directorate, Pacific Northwest National Laboratory, Richland, Washington 99352, USA*

[3]*Energy & Environment Directorate, Pacific Northwest National Laboratory, Richland, Washington 99352, USA*



A high sensitivity atom flux sensor based on atomic absorption spectroscopy has been designed and implemented to control electron beam evaporators and effusion cells in a molecular beam epitaxy system. Using a high-resolution spectrometer and a two-dimensional charge coupled device detector in a double-beam configuration, we employ either a non-resonant line or a resonant line with low cross section from the same hollow cathode lamp as the reference for nearly perfect background correction and baseline drift removal. This setup also significantly shortens the warm-up time needed compared to other sensor technologies and drastically reduces the noise coming from the surrounding environment. In addition, the high-resolution spectrometer allows the most sensitive resonant line to be isolated and used to provide excellent signal-to-noise ratio.


**Keywords:** Atomic absorption; MBE; Flux measurement; Baseline drift; Background correction


* U.S. patent pending.

a) Author to whom correspondence should be addressed. Electronic mail: yingge.du@pnnl.gov

b) Author to whom correspondence should be addressed. Electronic mail: scott.chambers@pnnl.gov






Molecular beam epitaxy (MBE) has been the method of choice for deposition of single crystalline thin films and heterostructures of the highest quality since its invention in the late 1960s. Compared with other deposition techniques, MBE enables the highest level of source purity, flexibility, and control over structural perfection and layer thickness. However, its widespread use has been hampered by difficulties associated with stoichiometry control.[1, 2] The difficulty lies in the fact that in MBE, each individual element is generated as a separate atom beam which must be delivered to the substrate with a precisely controlled flux in order to yield the desired stoichiometry and growth rate. These fluxes can change over time for many reasons, including shutter transients, depletion of source materials, and secondary phase formation on the surface of the hot metal source charges during deposition in reactive gas ambients (i.e., oxide formation during deposition in oxygen). As a result, element specific, real-time flux sensing and control at the sub-monolayer-per-minute level is critical for scalable and reproducible film synthesis.[3, 4]

Optical flux sensors based on the atomic absorption (AA) principle have been developed and employed to monitor atom fluxes in the past.[4-11] By passing a light beam matched to the absorption energy of the element of interest through the atomic beam, one can detect the intensity attenuation resulting from resonant atomic absorption. The attenuation is proportional to the atom density within the irradiated volume, and can be used to calibrate the atom flux and determine the deposition rate. AA flux sensors are element specific and non-protrusive within the deposition chamber, and can be highly sensitive. For maximum flexibility and broad applicability, hollow cathode lamps (HCL) are still the most widely used light sources, although diode lasers have also been used for certain elements.[3, 12, 13] Most of the existing AA-based flux sensor designs use a monochromator or bandpass filter (5 – 10 nm window) with a wide slit or





bandpass to select the wavelength of interest, and a photomultiplier tube (PMT) or a charge coupled device (CCD) for detection.[4, 6, 7, 9] The disadvantages of using a wide slit or wide bandpass width are poor sensitivity due to the possibility of inclusion of multiple lines in the detected signal, and the presence of non-linearities in the calibration curves.[14, 15] Many existing AA-based flux sensor designs utilize a normalization scheme in which the absorption is measured in relation to the total HCL output. However, instabilities in the HCL output result in non-negligible baseline drift even when there is no atom flux in the optical path, particularly during the first hour of operation in constant-current mode.[14] Additional sources of noise or baseline drift include unintentional coating or other real-time changes to hardware in the optical path that alter the transmission, and slight optical misalignment or motion in the individual optical components that produce spurious changes in light intensity. A double-beam setup is often adopted to compensate for long-term changes in the source intensity and detector sensitivity. Frequent baseline re-calibrations while blocking the atom beam can circumvent baseline stability issues,[9] but such steps can be difficult to implement and can interfere with deposition processes, especially automated ones. In one of the first commercially available AA-based flux sensors, a double-beam optical configuration was combined with the common optical path for automatic correction of transmission (COPACT) geometry, in which a xenon flash lamp was used as a reference. This scheme was shown to improve long-term stability in the baseline, enabling continuous detection, although a manual re-zero is typically required before each experiment.[4] The use of a xenon flash lamp, however, introduces another potential source of instability, and requires a portion of the detection time, thereby lowering the overall signal-to-noise ratio and measurement frequency. For Si, Cu, and Ge, the detection and control limits of an AA sensor with the COPACT configuration were found to be in the range of ~0.1Å/s.[4] A less





frequently discussed approach in traditional AA spectroscopy is the two-line background correction, in which the total absorbance is measured on both a resonant line and either a non-resonant or a weak resonant line, and the difference is taken.[16-18] This approach is often overlooked because of difficulty identifying suitable non-resonant lines and the complexity inherent in using this approach with existing designs. To the best of our knowledge, this approach has not been reported for background correction in AA-based atom flux sensors.

In this work, we have combined a high-resolution spectrometer and high-precision 2-D digital CCD camera in a simple double-beam configuration for accurate flux sensing and control during MBE deposition. The use of a high-resolution spectrometer allows selection of only the analytical peak of interest, resulting in optimal signal response and enhanced signal-to-noise (S/N) ratio.[14] The use of a thermoelectrically cooled CCD camera minimizes dark current and offers enhanced S/N ratio. Most importantly, we show that the simultaneous detection of the strongest absorbed resonant analytical line, and other weakly resonant or non-resonant emission lines, enables superior background correction and baseline drift removal, both of which are critical for long-term stability. This self-correcting scheme also offers other unique advantages, including a reduced system warm-up time and less sensitivity to external disturbances, such as slight disturbance to the fiber optic cable during measurement.

Our AA system, as shown in Fig. 1, is coupled to the deposition process through two fused silica viewports using customized mounting assemblies to provide non-intrusive monitoring and control. The HCLs are operated in constant-current mode. A 70/30 optical beam splitter is used to separate the HCL light into two beams, defined as the sampling and reference channels, respectively. The sampling channel beam with higher intensity is focused onto a





multimode (1 mm core diameter) optical fiber (denoted as the emitting fiber optic), passes

through a collimator coupled with the viewport and MBE chamber, and is collected by a UV-VIS

bifurcated optical fiber bundle. The reference channel is collected directly from the beam splitter

by the reference leg of the same bifurcated fiber bundle. With each leg consisting of 7 subunits

of 200 μm diameter fibers in a "1+6" configuration on a SMA-905 connector (cross sections for

sampling leg and reference leg are labeled as "1" and "2", respectively, in Fig. 1), the bifurcated

fibers carry the beams and run them into the spectrometer with a 10 mm ferrule on the common

end in a "7+7" configuration (cross section is labeled as "3" in Fig. 1). The spectrometer

(Princeton Instrument (PI), Acton SP 2558) disperses the two beams onto a thermoelectrically

cooled, high sensitivity 2-D CCD camera (PI Acton PIXIS:400F, 1340x400, 20x20 μm size).

When the optical fiber arrays are vertically aligned at the entrance slit, the resulting image on the

CCD is spatially resolved and can be separated into two regions: the sampling signal and the

reference signal. The sampling and reference spectra are dispersed by wavelength along the 1340

pixels of the CCD camera, using a standard grating (1200 G/mm with 300 nm blaze wavelength),

along with a motorized mirror drive. Multiple emission lines in the pm width range can be

observed as isolated dot arrays with the resolution determined by the size of entrance slit, in this

case, 200 μm, the diameter of the fiber. The outputs of the digital CCD camera are read in

imaging mode and then vertically summed within the sampling and reference regions to yield

two spectra. The intensity from one irradiation line is then calculated by horizontally integrating

over the designated pixels. We have developed customized software to process the data, analyze

the absorption signal, and provide feedback control to the power supply via a proportional-

integral-derivative (PID) algorithm, which can effectively regulate the atom flux.





At very low flux and under monochromatic radiation, light absorption follows the Beer-Lambert law:[14]

$$A(h\nu, t) = log_{10}(\frac{I_{in}}{I_{out}}) \propto n(t)Lk(h\nu) \qquad (1)$$

Here $A$ is the absorbance, $h$ is the Plank's constant, $\nu$ is the light frequency, $t$ is time, $I_{in}$ is the light intensity incident on the atomic beam, $I_{out}$ is the light intensity after passing through the atomic beam, $n$ is atomic density therein, $L$ is the length of the absorbing path, and $k$ is the absorption coefficient. In a double beam configuration, $I_{out}$ is directly measured as the sampling intensity, $I_{sample}$. $I_{in}$ is the sampling light intensity before entering the MBE chamber and is calculated by:

$$I_{in}(h\nu, t) = I_{ref}(h\nu, t) * (\frac{I_{sample_o}}{I_{ref_o}}) \qquad (2)$$

Here $I_{ref}$ is the reference channel intensity, $I_{ref_o}$ and $I_{sample_o}$ are reference and sampling intensities measured at a specific time, typically right after the system warm-up, when the atomic flux is 0. Combining equation (1) and (2) yields

$$A(h\nu, t) = log_{10}(c * \frac{I_{ref}}{I_{sample}}) \qquad (3)$$

Here $c = I_{sample_o}/I_{ref_o}$, and is refreshed whenever a "Re-zero" order is given. One assumption implicit in eq. 2 is that the ratio of $I_{sample}$ to $I_{ref_o}$, determined by the coefficients of the beam splitter and optical path, does not vary with time. However, during the warm-up phase, the beam splitter is under HCL illumination and thus temperature changes may lead to mechanical misalignment and/or initial changes in reflection and transmission coefficient until thermal





equilibrium is reached. In addition, for a HCL in the warm-up phase, the light intensity as well as the plasma position change inside the HCL.[14] The latter affects the optical alignment between the HCL and beam splitter, and thus cannot be mitigated by the use of a xenon flash lamp or deuterium lamp as a reference. As a result, a warm-up period of ~1 hr is routinely required. However, if one can simultaneously detect a nearby emission line that originates from the same HCL and passes through the same beam splitter and optical paths as the analytical resonant line, the effect of the time dependent changes mentioned can be monitored and corrected for.

In our system we use a non-resonant line, or in certain circumstances a resonant line with low absorbance, from the HCL for baseline drift removal. Because the atoms in the beam are in their ground electronic states, only photons with energies equal to the energy difference between the ground state and some excited state are absorbed. As a result, emission lines associated with transitions involving two excited states, and inert gas emission lines, generated in the HCL will not be absorbed. Theoretically, the absorbance should be zero throughout the experiment for all non-resonant lines. However, in the actual experimental configuration, the apparent absorbance of the non-resonant line ($A_{non-res}$) tends to slowly and monotonically increase/decrease or "drift" over time, in a way that is proportional to the baseline drift. Thus, $A_{non-res}$ is a direct measure of baseline drift and provides an ideal signal for real-time correction. The true absorbance after correction is given by $A = A_{res} - A_{non-res}$, where $A_{res}$ is the absorbance of a resonant line. In addition, because the non-resonant and resonant lines originate from the same HCL and go through the same optical path, any background absorption associated with viewport coatings or optical misalignment will affect both $A_{res}$ and $A_{non-res}$ in the same way, and thus will be removed.





Fig. 2(a) shows the emission spectra from a Cr HCL (PerkinElmer, Atomax, 12 mA) collected on the 2-D CCD after the HCL has been operating for 20 min. There are three strong resonant lines, labeled as *A1*, *A2*, and *A3*, fall at 357.9, 359.3, and 360.5 nm, respectively. [19] In addition, there is a non-resonant line at 352.0 nm, labeled *A4*. Fig 2b shows the absorbance signal using the double-beam technique when the sensor is used to monitor the Cr flux coming from an effusion cell held at $1410^{o}$C. For the first 16 minutes of the measurement, a closed source shutter blocks the beam. The shutter is opened at 16 min and strong absorbance signals are observed for all resonant lines whereas little change is observed in the non-resonant line. It is clear that *A1* is the resonant line with the highest absorption coefficient and thus offers the best sensitivity. When the shutter is closed at 24 min, the absorbance signals for *A1*, *A2*, and *A3* all drop down to the level of *A4*, but all four lines display a non-zero reading. This baseline drift is spurious and detrimental for quantitative analysis and process control. However, *A4* can be effectively used to correct for this baseline drift. In Fig. 2(c), (*A1 - A4*) is plotted to show absorbance vs. time with our self-corrected double-beam configuration. The self-corrected absorbance (*A1 - A4*) thus gives an accurate measure of Cr atom flux, replete with removal of a shutter transient upon initial shutter opening, and returns to zero absorbance when the shutter is closed. It should be noted that if a non-resonant line is not readily available, a resonant line with a lower absorbance (i.e. a smaller AA cross section) can be used for background correction. For example, if (*A1-A2*) or (*A1-A3*) in Fig. 2(b) were plotted, the background correction would be very similar to that seen in Fig. 2(c). One disadvantage in using a resonant line for this purpose is that the signal-to-noise ratio will be lower as the magnitude of the absorbance is smaller. It should also be noted that if sufficient warm-up time is allowed (e.g, 60 min rather than 20 min), the baseline drift detected over a 30 min range can be further reduced, but not completely





eliminated for the direct double-beam configuration (Fig. 2(b)). By using the non-resonant line *A4* for self-correction, this magnitude of drift, which is inherent in the double-beam setup, is effectively removed without extensive and painstaking mechanical and thermal stabilization, and thus, a much shorter warm-up time is needed.

In MBE, it is often critically important to control individual fluxes at an extremely slow, reproducible, and known rate. By implementing a proportional-integral-derivative (PID) control loop, the AA software can supply the signal utilized to regulate the power supply of an electron-beam or Knudsen cell evaporation source to maintain constant absorbance during evaporation. For small absorbance, the deposition rate on the substrate and the absorbance have a linear relationship. The deposition rate can be experimentally determined by placing a quartz crystal oscillator (QCO) at either the sample growth position, or alongside the sample manipulator.  In the latter case, a tooling factor is needed to compensate for the difference in position between substrate and QCO sensor. For materials which grow in a controlled layer-by-layer fashion, it is also possible to accurately measure the deposition rate by monitoring the reflection high-energy electron diffraction (RHEED) intensity oscillations during MBE growth.[20]

Utilizing both a QCO sensor and the RHEED intensity oscillations from epitaxial single crystalline α-$Cr_2O_3$/α-$Al_2O_3$(0001) film growth  to provide a time-dependent atomic flux rate comparison, the AA sensor was used to monitor and control the evaporation rate of Cr from an electron beam evaporator. During the 90 min of deposition, the α-$Al_2O_3$ substrate was heated to 500$^o$C and exposed to oxygen plasma at a pressure of 1x10$^{-5}$ Torr to facilitate complete oxidation of Cr atoms. Fig. 3(a) shows a screen shot of the AA software, revealing that the absorbance is maintained at 0.10.  The QCO was programmed with the input parameters for $Cr_2O_3$ growth, but





the rate shown is not accurate in an absolute sense because the QCO was located off to the side and an accurate tooling factor was not pre-determined prior to deposition. In addition, considerable time (tens of minutes) is needed for a QCO sensor to reach thermal equilibrium and provide consistent readings once the source shutter is opened.[21] To this end, we allowed 40 min for the QCO to reach thermal equilibrium and thus only showed the rate during the last 50 min of deposition in Fig. 3(b). The deposition rate was measured to be 0.0357 Å/sec at the beginning and 0.0363 Å/sec at the end of growth. The flux increased slightly over a 50 min period by 0.0006 Å/sec (+1.7%), as determined from a linear least-squares fit shown as the line in Fig. 3(b). This flux change resulted in a change in the growth rate of the film of a similar magnitude (+1.5%), as measured by RHEED intensity oscillations and discussed in the Supplemental Material at (URL to be inserted ). The fact that such a low growth rate can be held steady over the course of 90 min shows that the AA system described here allows unprecedented sensing and control of the atom beam flux.

We have tested the sensor on other elements as well, including La, Fe, Zr, and Ta. The self-correction scheme has proven to provide a more stable baseline and better signal-to-noise ratio than what can be achieved with existing technology in all cases. As with other designs, the sensitivity for each element is determined primarily by the stability of the lamp, the intensity of the analytical emission line, and the cross-section for absorbance.[14] Once an absorbance-deposition rate calibration curve is established, the AA is easier to use and more practical than a QCO, particularly for e-beam evaporations where the thermal load can cause instability and spurious readings in a QCO. A comparison for Zr evaporation is shown in the Supplemental Material. Over an extended period of time, the actual deposition rate may deviate from the





calibration curve for reasons such as a change of shape or height of the metal charge, and a re-calibration should be performed.

By combining a high-resolution spectrometer with a 2-D CCD camera, we have implemented a self-corrected, double-beam AA sensor to monitor and control atom beam fluxes in a MBE chamber. A non-resonant line or a resonant line with low absorbance near the analytical line from the same HCL has been utilized for effective background correction and baseline drift removal. The sensor exhibits superior sensitivity with a stable baseline. It should be noted that this configuration can be used for simultaneous, multi-element, and multi-channel detection when equipped with appropriate multi-element HCLs as light sources. Our double-beam, baseline and background self-correction scheme is not only useful for flux sensing in thin-film growth, but also for the overall development of atomic absorption spectrometry as an analytical technique.

### Acknowledgements

This research was supported by the U.S. Department of Energy (DOE), Office of Environmental and Biological Research under a Major Item of Equipment grant for the development of a state-of-the-art oxide MBE system, and the Office of Basic Energy Sciences, Division of Materials Science and Engineering under Award 10122. The work was performed at the W. R. Wiley Environmental Molecular Sciences Laboratory, a DOE User Facility sponsored by the Office of Biological and Environmental Research. The authors would like to thank Tiffany Kaspar and Hongfei Wang for many insightful discussions.





## References:


1. L. Qiao, K. H. L. Zhang, M. E. Bowden, T. Varga, V. Shutthanandan, R. Colby, Y. Du, B. Kabius, P. V. Sushko, M. D. Biegalski and S. A. Chambers, Adv. Funct. Mater. **23** (23), 2953-2963 (2013).
2. G. L. Olson, J. A. Roth, P. D. Brewer, R. D. Rajavel, D. M. Jamba, J. E. Jensen and B. Johs, J. Electron. Mater. **28** (6), 749-755 (1999).
3. C. Buzea and K. Robbie, Rep. Prog. Phys. **68** (2), 385-409 (2005).
4. C. Lu and Y. Guan, J. Vac. Sci. Technol., A **13** (3), 1797-1801 (1995).
5. A. W. Jackson, P. R. Pinsukanjana, A. C. Gossard and L. A. Coldren, Ieee Journal of Selected Topics in Quantum Electronics **3** (3), 836-844 (1997).
6. S. A. Chalmers, K. P. Killeen and E. D. Jones, Appl. Phys. Lett. **65** (1), 4-6 (1994).
7. S. A. Chalmers and K. P. Killeen, Appl. Phys. Lett. **63** (23), 3131-3133 (1993).
8. J. A. Silver, Appl. Opt. **31** (6), 707-717 (1992).
9. M. E. Klausmeierbrown, J. N. Eckstein, I. Bozovic and G. F. Virshup, Appl. Phys. Lett. **60** (5), 657-659 (1992).
10. T. Y. Kometani and W. Wiegmann, J. Vac. Sci. Technol. **12** (4), 933-936 (1975).
11. Y. Kasai and S. Sakai, Rev. Sci. Instrum. **68** (7), 2850-2855 (1997).
12. O. M. Marago, B. Fazio, P. G. Gucciardi and E. Arimondo, Appl. Phys. B **77** (8), 809-815 (2003).
13. W. Z. Wang, R. H. Hammond, M. M. Fejer, C. H. Ahn, M. R. Beasley, M. D. Levenson and M. L. Bortz, Appl. Phys. Lett. **67** (10), 1375-1377 (1995).
14. B. Welz and M. Sperling, in *Atomic Absorption Spectrometry* (Wiley-VCH Verlag GmbH, 2007), pp. 63-147.
15. L. De Galan and G. F. Samaey, Spectrochim. Acta Part B **24** (12), 679-683 (1969).
16. T. Ashino, S. Morimoto and K. Wagatsuma, Anal. Sci. **26** (12), 1301-1304 (2010).
17. M. T. C. de Loos-Vollebregt, in *Encyclopedia of Analytical Chemistry* (John Wiley & Sons, Ltd, 2006).
18. S. L. Tong and K. S. Chin, Spectrochim. Acta, Part B **49** (5), 459-467 (1994).
19. C. C. Kiess, J. Res. Nat. Bur. Stand. **51** (5), 247-305 (1953).
20. S. A. Chambers, Y. Liang and Y. Gao, Phys. Rev. B **61** (19), 13223-13229 (2000).
21. Y. S. Kim, N. Bansal, C. Chaparro, H. Gross and S. Oh, J. Vac. Sci. Technol. A **28** (2), 271-276 (2010).






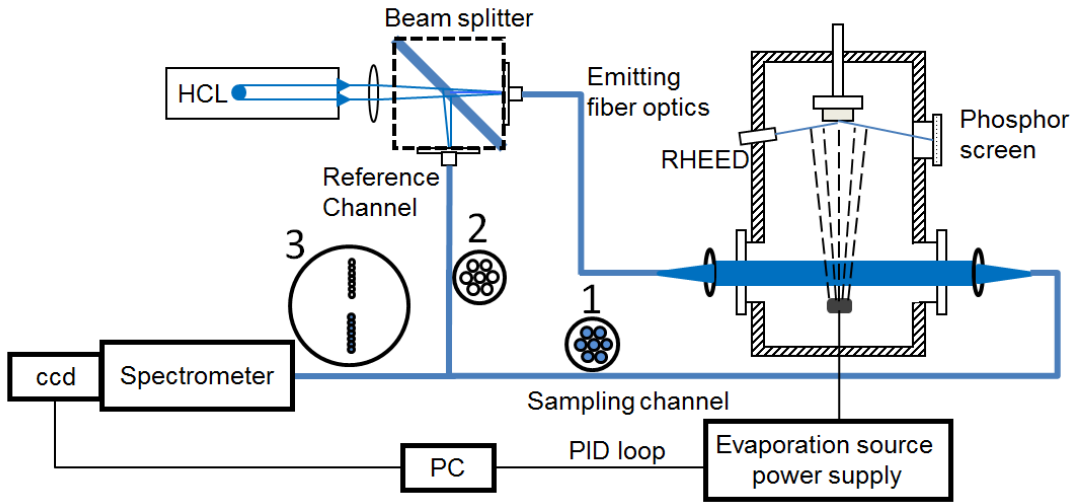

FIG. 1. Overall schematic configuration for a self-corrected, double-beam atom beam flux sensor.





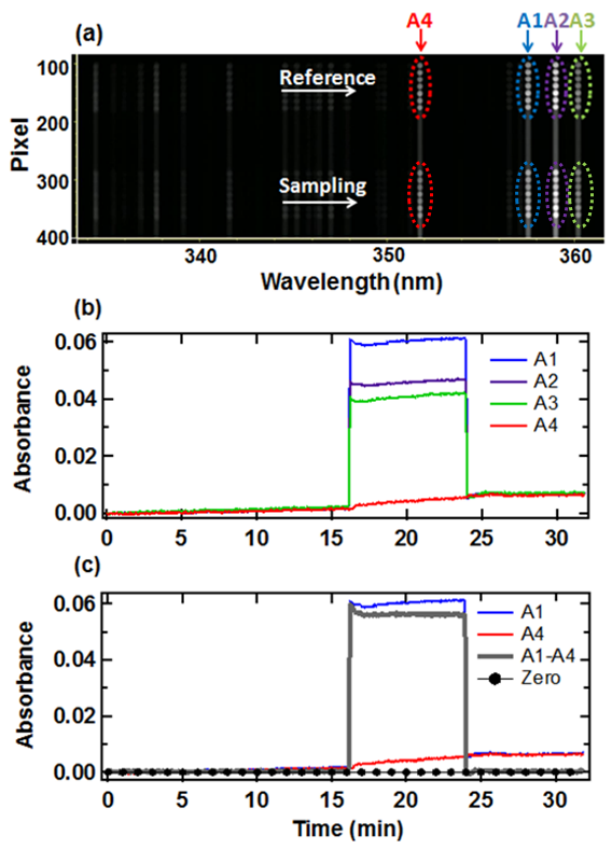

FIG. 2. (a) CCD camera image of the emission spectrum from a Cr hollow cathode lamp; (b) Cr absorbance measured with the standard double-beam configuration; (c) Cr absorbance measured using a self-corrected double-beam approach.





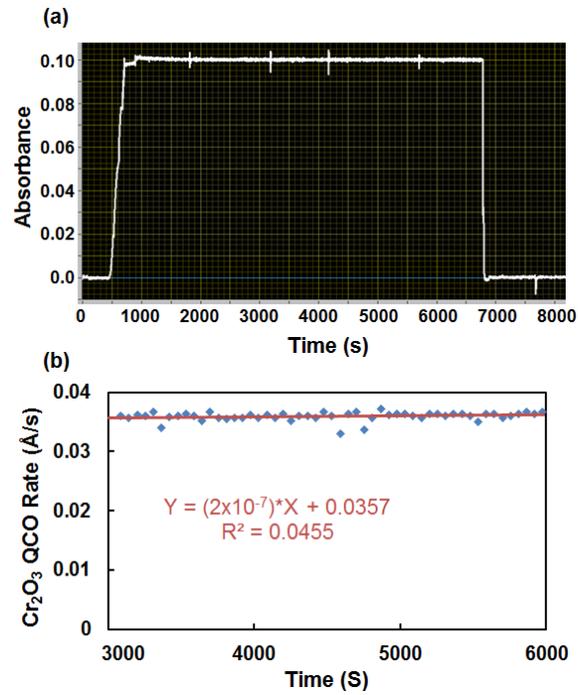

FIG. 3. (a) Experimental trace for the atomic absorption controlled electron beam evaporation of Cr; (b) Cr flux measurement by a quartz crystal oscillator at the side of sample manipulator during the last 3000 seconds of the deposition recorded in panel (a).





# Supplemental Material for

# Self-corrected sensors based on atomic absorption spectroscopy for atom flux measurements in molecular beam epitaxy


Y. Du, T. C. Droubay, A. V. Liyu, G. Li, and S. A. Chambers


The RHEED patterns for the sapphire substrate and the $Cr_2O_3$ film, along with the specular RHEED beam intensity oscillations were recorded during the epitaxial film growth described in the Letter. Fig. S1(a) shows the RHEED patterns for the $\alpha$-$Al_2O_3$(0001) substrate and the $Cr_2O_3$(0001) film along two different azimuthal directions. The sharp, streaky patterns for the $Cr_2O_3$ film demonstrate the epitaxial orientation and excellent structural quality of the film. More importantly, the RHEED intensity oscillations (Fig. S1(b)) for the specular beam (dotted oval in Fig. S1(a)) reveal that the film grows in a layer-by-layer fashion. An average $Cr_2O_3$ growth rate can be determined from the period of oscillations (~48 sec) along with the interlayer spacing for $Cr_2O_3$(0001) (~2.3Å). Reasonably strong oscillations appeared shortly after the deposition started and persisted throughout the deposition process (90 min). An expanded view shown in the inset for Fig. S1(b) reveals that the period is 48.1 sec (corresponding to a film growth rate of 0.0478 Å $Cr_2O_3$/s) at the beginning. The period dropped to 47.4 sec by the end of growth. This change corresponds to a +1.5% increase in actual growth rate, consistent with the QCO measurement of the Cr flux change discussed in the Letter. Over 90 min, a total of 113 oscillations, corresponding to 113 ML, were observed, giving a total thickness of 26.0 nm.

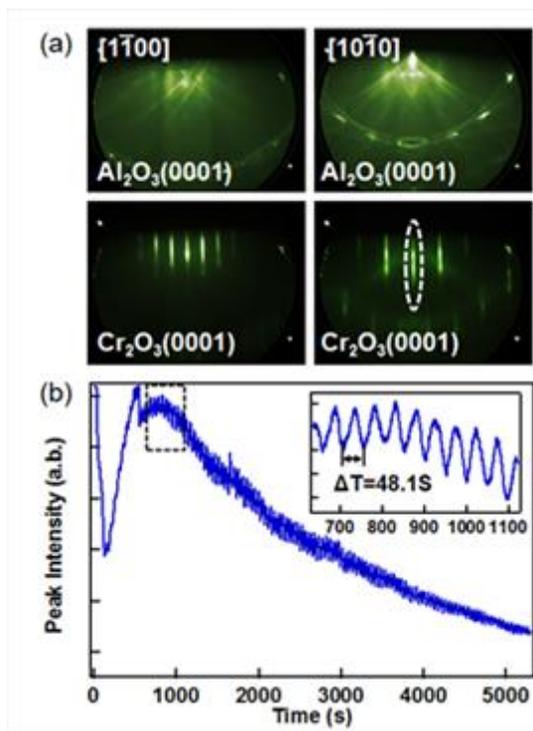

FIG. S1. (a) RHEED patterns for an $\alpha$-$Al_2O_3$(0001) substrate and a $\alpha$-$Cr_2O_3$(0001) film along different low-symmetry directions. (b) RHEED intensity oscillations for the specular beam intensity.

In the Cr data we used for the paper, the absorbance was set to 0.1 and the standard deviation resulting from the shot noise measured over several tens of seconds was ~0.0002, which gives an accuracy of ~0.2% in controlling the absorbance signal. However, the AA system clearly did not detect the 1.7% increase in flux recorded by the QCO and RHEED oscillations. There are two possible reasons for this outcome. One is that although highly effective, the self-correction scheme employed here cannot perfectly remove the baseline drift or other background changes of unknown origin. The second is that the change in flux occurred as a result of the Cr metal charge changing shape, resulting in an increase in the cross sectional area of the atom beam, but no change in the total evaporation rate. In this scenario, the AA probe would not be as sensitive as either the QCO or the RHEED oscillations because the optical beam





passes through the entire atom beam along the light propagation direction, thus being inherently insensitive to changes in the cross sectional area of the beam. In contrast, both the QCO readings and RHEED oscillations on the film surface, being measured over small cross sectional dimensions of the atom beam, would be more sensitive to changes in the total flux. There is no easy way to distinguish between these two possibilities. It will take considerably more testing to determine ultimate sensitivity, and such tests are beyond the scope of this paper.

We use e-beam evaporation of Zr as a second example to compare the performance of our AA with that of a QCO. Fig. S2a shows a screen shot of the AA measurement/control interface. At approximately 5850s, the e-gun was manually adjusted to introduce small flux changes, and the AA was used only to monitor the associated changes in absorbance. From 5900s, the AA software took control of the power supply and maintained an absorbance at 0.019 and 0.004 respectively as indicated by the two rectangular

boxes. The QCO, which was placed at the normal substrate position, also recorded the growth rate during the entire time as shown in the QCO screenshot (Fig. S2b). It is clear that until the end of the first part of the AA controlled evaporation (absorbance = 0.019) at approximately 6100s, the measurements from both the AA and QCO are in very good agreement, including the detailed fluctuations. However, when the emission current of the e-gun was adjusted to maintain an absorbance at 0.004, the QCO reading became highly unstable, with an artificial increase followed by zero readings shortly thereafter. This occurred because the QCO is very sensitive to temperature and radiant heating by the hot metal charge changed significantly during these adjustments. When the e-gun was finally ramped down to zero after 6300s, the QCO again showed a spurious positive reading due to thermal shock.

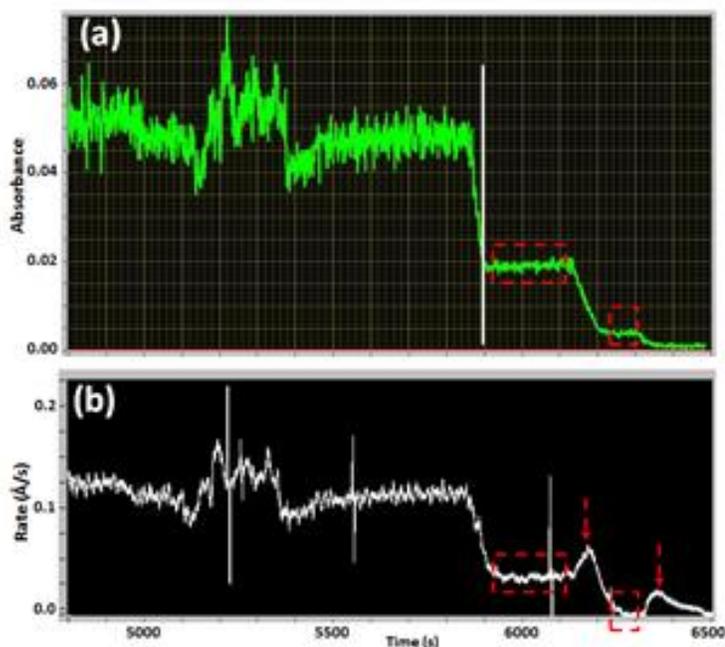

FIG. S2. Screen shots of the a) AA monitor/control interface and b) QCO measurement interface during the electron beam evaporation of Zr. The white line in a) separates the manual mode (left) from AA feedback controlled mode (right). Regions of AA control are indicated by red rectangles. Spurious readings are shown in the QCO measurement starting at the arrows due to thermal impact.